\documentstyle[aps,twocolumn,epsf]{revtex}

\newcommand{\be}{\begin{eqnarray}}
\newcommand{\ee}{\end{eqnarray}}

\newcommand{\la}{\langle}
\newcommand{\ra}{\rangle}

\begin{document}

\title{Dimensional Reduction in High Temperature Gauge Theories}

\author{Saumen Datta and Sourendu Gupta}

\address{Department of Theoretical Physics, Tata Institute of Fundamental
         Research, Homi Bhabha Road, Mumbai 400005, India}

\date{\today}

\maketitle

\begin{abstract}
Screening masses in the equilibrium thermodynamics of gauge theories
were measured. The spectroscopy of these screening masses gives strong
evidence for dimensional reduction at $T\ge2T_c$. A perturbative
explanation of these masses is ruled out. The mass ratios in the high
temperature phase are consistent with those in the pure gauge theory
in three dimensions.
\end{abstract}

\pacs{}

\narrowtext


\section{Introduction}
\label{sec1}
 
Upto the present the only argument for dimensional reduction of
finite temperature field theories has been in terms of a mode
expansion of fields. Then it is argued that this mode expansion
may be truncated to the lowest Matsubara mode when considering
long distance physics \cite{feyn}. The consistency of this
perturbative argument has been checked by many recent works
\cite{keijo}.

Here I present a fully non-perturbative argument for dimensional
reduction \cite{paper}, involving a measurement of
the spectrum of screening masses in a equilibrium 3+1 dimensional
gauge theory. Degeneracies of the spectrum yield information on
the symmetry group of the spatial transfer matrix. This
turns out to be the rotation group in one lower dimension.

I also present one out of several arguments that this spectrum has
no perturbative explanation. The rest may be found in \cite{paper}.

Thermodynamics of field theories can be formulated in terms of a
spatial transfer matrix, $\cal T$, which propagates field
information from one point to another. On the lattice this is a
regulated object whose continuum limit has to be taken in the usual
way. The eigenvalues of $\cal T$ can be ordered ($\Lambda_0\le
\Lambda_1\le\cdots$). The free energy, and hence all thermodynamics
is entirely given by $\Lambda_0(T)$, where $T$ is the temperature.

The screening masses, $\mu$, are defined by the long distance
exponential fall of static current-current correlation functions
\be
   C_r(x)\equiv\la J_r(0) J_r(x)\ra\sim \exp(-\mu_r x),
\label{mu}\ee
where $r$ is a complete set of quantum numbers and $x$ a space-like
seperation. In terms of the eigenvalues of $\cal T$, $\mu_r=
\log(\Lambda_0/\Lambda_r)$, where the relevant eigenvalue depends on
the quantum number $r$. Thus, the screening masses contain more
information on the transfer matrix than bulk thermodynamics. In
particular, the degeneracies of $\mu_r$ tell us of the degeneracies
of $\Lambda_r$, and hence the effective symmetries of the transfer
matrix.

It helps intuition to note that the temporal transfer matrix of
a zero-temperature field theory is the exponential of the time
integral of the Hamiltonian of the theory. Hence its eigenstates
are the physical states of the theory and the eigenvalues are their
masses.

\section{Group Theory} 
\label{sec2}

In $d+1$ dimensions, a transfer matrix propagates information from
one $d$-dimensional slice of space-time to another. The transfer
matrix is invariant under any symmetry operation of this slice.

At zero temperature, an Euclidean field theory has $O(4)$
rotational symmetry. Hence the three dimensional slices relevant for
temporal and spatial transfer matrices are isomorphic.
Each such slice has $O(3)$ rotational symmetry. On the lattice this
breaks to the symmetry of a cube, $O_h$. This group includes the
parity $P$ with action $(x,y,z)\to(-x,-y,-z)$.

At finite temperature, $T$, the Euclidean time direction has extent
$1/T$, and hence is special. One consequence of this is well-known
to practitioners of perturbative finite-temperature field theory. It
is that 4-tensors have to be decomposed into spatial and temporal
(and various types of mixed) parts which transform differently. For
the classification of eigenstates of the spatial transfer matrix
the consequences are simple. The $O(3)$ symmetry of the $T=0$ theory
breaks down to the cylindrical symmetry, $\cal C$, consisting of all
operations in $O(3)$ which do not mix the time coordinate with $x$
and $y$. On the lattice, $O_h$ breaks to the dihedral group $D^4_h$.
The groups $\cal C$ and $D^4_h$ contain a parity $P$ with action
$(t,x,y)\to(-t,-x,-y)$, and the reflection $\sigma_z$ with action
$(t,x,y)\to(-t,x,y)$. The operators $P$ and $\sigma_z$ commute.

If dimensional reduction happens to be a good approximation to the
long distance physics, then the lowest eigenstates of the transfer
matrix can be classified into approximate multiplets of the two
dimensional rotation group, $O(2)$. On the lattice this would be
the group $C^4_v$. Both these contain the two dimensional parity
$\Pi$ with action $(x,y)\to(x,-y)$. The three $Z_2$ generators
are related by $\Pi=P\sigma_z$.

The real irreps of $O(2)$ can be obtained from the $O(3)$ angular
momentum irreps. There are two one dimensional irreps $0_+^P$ and
$0_-^P$ (the $J_z=0$ part of even and odd $J$ irreps) and a two
dimensional irreps for each distinct $\pm J_z$ projections of any
$O(3)$ irrep $J$.

The various groups are related by
\begin{equation}
 \matrix{
     O(3)=SO(3)\times Z_2(P) &\supset& O_h=O\times Z_2(P)\cr
     \bigcup & & \bigcup \cr\cr
     {\cal C}=O(2)\times Z_2(\sigma_z) &\supset&
                         D^4_h=D^4\times Z_2(P)\cr
     \bigcup & & \bigcup \cr\cr
     O(2) &\supset& C^4_v\cr
        }
\label{redux}\end{equation}
The irreps of the continuum groups can now be easily written.
The representation theory for $O_h$ in terms of loop operators was
first presented in \cite{alain}. The identification of $\cal C$ and
$D^4_h$ as the appropriate symmetries at $T\ne0$ was made in
\cite{old} where the representation theory of $D^4_h$ in terms of
gluon operators was given. Similar work was also done in \cite{yaffe}.

Full details of the reduction formul\ae{} can be found in \cite{paper}.
Here we only quote the results necessary to make sense of the numerical
data to be presented later. If dimensional reduction occurs, then
some representations of $D^4_h$ collapse pairwise into representations
of $C^4_v$. We write the results for the one-dimensional irreps of the
two groups---
\begin{eqnarray}
   A_1^{P,C},A_2^{-P,C}\;&\to&\;A^{\Pi=P,C},\cr
   B_1^{P,C},B_2^{-P,C}\;&\to&\;B^{\Pi=P,C}.
\label{d4htoc4}\end{eqnarray}
Here $C$ denotes the charge conjugation quantum number. If lattice
artifacts are small, then the splitting between the $B^{++}$ and
$B^{-+}$ irreps of $C^4_v$ vanishes.

\section{Dimensional Reduction}
\label{sec3}

In this section we present results from simulations of the $SU(3)$
gauge theory. The simulations were performed using a heat-bath
algorithm \cite{kphb}. Particulars of the runs are summarised in
Table \ref{tab1}. Correlation functions between loop operators were
measured in the long direction of the lattice, after averaging over
the transverse slice. Along this long direction we measured correlation
functions at seperations of more than $1/T$. Previous measurements at
these temperatures show that the gauge coupling $g>1$ \cite{spstr}, so
that $1/g^2T<1/gT<1/T$.

\begin{table}[bht]\begin{center}
  \begin{tabular}{|r|c|c|r|}  \hline
  $T$ & $\beta$ & Size & Statistics \\ 
  \hline
  $T_c$ & 5.7 & $4.8^2.16$ & 1000 + {\bf 5000} ($\times 50$) \\
  ${3\over2} T_c$
        & 5.9 & $4.8^2.16$ & 400 + {\bf 5000} ($\times 10$) \\
  $2 T_c$
        & 6.0 & $4.8^2.12$ & 400 + {\bf 5000} ($\times 10$) \\
        &     & $4.8^2.16$ & 400 + {\bf 10000} ($\times 10$) \\
        &     & $4.12^2.16$ & 400 + {\bf 10000} ($\times 10$) \\
  \hline
  \end{tabular}\end{center}
  \caption{Details of runs on $N_t=4$ lattices for $SU(3)$ pure gauge
     theory. The correlation functions are measured in the long direction
     which is always kept greater than $4/T$. The statistics is quoted
     as $\rm discarded+configurations\times seperation$.}
\label{tab1}\end{table}

We measured 95 operators for $A_1^{++}$, 80 for $B_1^{++}$, 55 for $A_2^{++}$
and 40 each for $A_1^{-+}$, $B_1^{-+}$ and $B_2^{++}$. For noise reduction we
used a variant of the usual techniques \cite{fuzz}. The lowest screening
mass in each channel was found by an algorithm \cite{vary} which implements
the following idea--- a linear combination of operators is taken in the
representation $r$,
\be
   {\cal O}^r\;=\;\sum_i \alpha_i {\hat O}_i^r,
\label{vary}\ee
(the index $i$ runs over all operators in the measured set) and the
coefficients $\alpha_i$ are varied to minimise the measured screening mass.

The results are given in Table \ref{tab2}. The measured values of $\mu/T$
in the $A_1^{++}$ and $A_2^{-+}$ irreps of $D^4_h$ should become degenerate
when dimensional reduction works. This happens at $T=2T_c$ but not for
$T\le 3T_c/2$. Also at $2T_c$, the screening masses in the irreps
$B_1^{-+}$ and $B_2^{++}$ of $D^4_h$ become degenerate, giving additional
evidence for dimensional reduction on the lattice.

Lattice artifacts are under reasonable control and the continuum physics
is not very far away. Evidence for this comes from the splitting between
the screening masses of the $B^{++}$ and $B^{-+}$ irreps of $C^4_v$,
which correspond to the same irrep of the continuum group. This splitting
is less than 14\% of the average of these two screening masses.

The screening mass ratios compare well with the glueball mass ratios in
$SU(3)$ pure gauge theory in three dimensions. We find
\be
   \matrix{
      {\displaystyle\mu(A^{++})\over\displaystyle\mu(B^{++})}
                                                 \;=\;0.54\pm0.02,\cr
      {\displaystyle\mu(B^{++})\over\displaystyle\mu(A^{-+})}
                                                 \;=\;0.76\pm0.02.\cr}
\label{rat}\ee
In the 3-d $SU(3)$ pure gauge theory these ratios are $0.60$ and $0.78$
respectively \cite{teper}.

We have also made measurements in the $SU(2)$ gauge theory \cite{paper}.
We find dimensional reduction at $2T_c$ in that the screening masses are
organised by $C^4_v$. In $SU(2)$ the lattice artifacts are larger, and a
comparison with pure gauge theory is still not possible. Finite-size
effects are currently under investigation.

\begin{table}[thb]\begin{center}
  \begin{tabular}{|c|c|c|l|l|l|}  \hline
  $O(2)$ & $C^4_v$ & $D^4_h$ & \multicolumn{3}{c|}{$\mu/T$} \\
  \cline{4-6} 
  & & & $T_c$ & ${3\over2}T_c$ & $2T_c$ \\
  \hline
  $O^+_+$ & $A^{++}$ & $A_1^{++}$ &
            $3.4\pm0.4$ & $2.56\pm0.04$ & $2.60\pm0.04$ \\
  $O^+_+$ & $A^{++}$ & $A_2^{-+}$ &
                      - & $3.2 \pm0.1 $ & $2.8 \pm0.2 $ \\
  $O^+_-$ & $A^{-+}$ & $A_1^{-+}$ &
                      - & $6.3 \pm0.1 $ & $6.3 \pm0.2 $ \\
  $2^+  $ & $B^{++}$ & $B_1^{++}$ &
            $4.9\pm0.3$ & $4.9 \pm0.3 $ & $4.8 \pm0.1 $ \\
  $2^+  $ & $B^{-+}$ & $B_1^{-+}$ &
                      - &             - & $5.5 \pm0.3 $ \\
  $2^+  $ & $B^{-+}$ & $B_2^{++}$ &
                      - & $5.1 \pm0.3 $ & $5.6 \pm0.4 $ \\
  \hline
  \end{tabular}\end{center}
  \caption{Screening masses in various irreps of $D^4_h$. The pattern
     of degeneracies follows closely the organisation into irreps of
     $C^4_v$. The organisation into irreps of the continuum dimensionally
     reduced group $O(2)$ is still not very good.}
\label{tab2}\end{table}

\section{Magnetic masses}
\label{sec4}

One interesting question is whether the data on screening masses can be
interpreted in terms of two basic masses--- the Debye screening mass,
$M_D$ and the magnetic mass, $M_m$. Complete details are given in
\cite{paper}.

In order to do this we construct a gluon field operator associated with
each link of the lattice. It must be an element of the Lie algebra of the
gauge group, and when exponentiated, must give the group element associated
with the link. The representations turn out to be gauge independent,
although the field values are not. With this construction in hand, we can
go on to construct multigluon operators. By matching the representation
content of these with the measured correlation functions we will be in a
position to test perturbation theory.

The gluons have $C=-1$. Hence all $C=1$ irreps are obtained from exchange
of an even number of gluons and all $C=-1$ from an odd number. Gauge-invariant
zero-momentum two gluon operators---
\be
   O_{\mu\nu}\;=\;\sum_k Tr[G_\mu(k) G_\nu(-k)]
\label{2gl}\ee
have $P=1$ after imposing symmetry under exchange of the two gluons. $P=-1$
and $C=1$ irreps can only be obtained by exchange of at least four gluons.

If we associate with each gluon some mass, and assume a perturbative
dispersion relation for the gluons, then the observed systematic degeneracy
of $P=1$ and $P=-1$ states obviously implies that one of the following
conditions is false---
\begin{enumerate}
\item
   The dispersion relations are perturbative.
\item
   The gluons can be assigned a non-vanishing mass.
\end{enumerate}
Since the degeneracy of opposite parity states is required for dimensional
reduction, it is not consistent with a perturbative spectrum of screening
masses.

\section{Summary}
\label{sec5}

We have studied gauge invariant screening masses in pure gauge theories
at high temperatures. We have found evidence of dimensional reduction in
the spectrum at $T=2T_c$, but not at lower temperature. In the $SU(3)$
theory, the screening mass spectrum agrees with the glueball mass
spectrum of 3-d pure gauge $SU(3)$ theory. Group theoretical considerations
show that dimensional reduction is not consistent with a perturbative
spectrum of screening masses.


\end{document}